# Dissociation of Nucleon and Heavy-Baryon in an Anisotropic Hot and Dense QCD Media Using Nikiforov-Uvarov Method

M. Abu-Shady[1*] and A. N. Ikot[2**]

[1]Department of Applied Mathematics, Faculty of Science, Menoufia University, Egypt

[2]Department of Physics, University of Port Harcourt, Nigeria

**Abstract**

By using the Nikiforov-Uvarov method, the hyper-radial Schrödinger equation is analytically solved, in which the real modified potential is employed at finite temperature and baryon chemical potential. The eigenvalue of energy and corresponding wave function are obtained in the isotropic and anisotropic media in the hot and dense media. The present results show that the binding energy of nucleon and some heavy baryon decrease strongly in hot medium and decreases slightly with increasing baryon chemical potential. In addition, the binding energy for each baryon is more bound in an anisotropic medium in comparison with its value in an isotropic medium. The dissociation of temperature of each baryon is above a critical temperature and it increases in the anisotropic medium. The dissociation of temperature is slightly decreased in the hot medium when the baryon of chemical potential is considered. A comparison is studied with the available studies. We conclude that the present study provides a good description of the nucleon and some heavy baryon in the hot and dense media in the isotropic and anisotropic systems.

**corresponding author:ndemikotphysics@gmail.com

# 1-Introduction

There is a great interest in studying quarkonia at finite temperature or /and finite baryon chemical potential which is one of the basic ingredient tools for understanding the status of quark-gluon plasma (QGP) [1-13]. This interest was firstly introduced by Matsui and Satz [14]. They discussed that the quarkonium dissociation is above a critical temperature when the color screening in the medium is accrued, which indicated as a signal quark-gluon plasma formation. The ultra-relativistic heavy-ion collision (URHIC) is used to explore the quark-gluon plasma (QGP) as in Refs. [15-17]. Most of these studies found that dissociation of charmonium and bottomonium are above a critical temperature. On the same hand, some of these studies are extended to include the effect of finite baryon chemical potential, in which the binding energy is dropped to lower values by increasing baryon chemical potential. In addition, the color-screening effect at finite temperature and chemical potential was studied in a thermodynamics field approach as in Refs. [18, 19] in which the phenomenological potential model [20] and an error-function-type confined force with a color-screened Coulomb-type potential were used. Lattice QCD had also been used to study the color screening in the heavy-quark potential at finite density with Wilson fermions [21]. Kakade and Patra [22] have investigated the quarkonium dissociation at lower temperatures and higher baryonic chemical potential by correcting both the perturbative and non-perturbative approaches of the Cornell potential through the dielectric permittivity in an isotropic medium.

In recent works, the anisotropic medium plays an important role that found the nuclei moves to each other when the collision has moved with ultra relativistic speed which indicated that there is a spatial anisotropy in the system. Therefore, the system requires thermalizing and isotropizing, which the anisotropic pressure gradient is expected in all directions. So it is important to include momentum anisotropy in the analysis {see Ref. [23] and references therein}. Also, the heavy-quark potential is extended to a local anisotropy in the momentum space [24]. In Refs. [25-29], the quarkonium properties in an anisotropic medium are investigated due to the presence of external fields.

There are few works in the studying of dissociation of nucleon and heavy- baryon in hot and dense media. The study of three bound states is a difficult task due to solving nonrelativistic quark model for three bound states. In Ref. [30], the authors studied the nucleon dissociation in a hot quantum chromodynamic medium using Gaussian expansion method to solve Schrödinger equation and they found that the binding energy of nucleon decreases with increasing temperature. In addition, the nucleon state dissociates above the critical temperature. On the other hand, the nucleon properties in hot and dense media using the chiral quark models has been investigated (see Refs. [31-36]).

In the present work, the dissociation of nucleon and heavy-baryon at finite temperature and chemical potential in the isotropic and anisotropic media is investigated. The Nikiforov-Uvarov (NU) method is used to obtain the analytic solution of the hyper-radial Schrödinger equation. The previous works are not considered the anisotropic medium. Thus, the present work is a unified treatment of dissociation of nucleon and heavy-baryon in the hot and dense media in an anisotropic medium using NU method.

The paper is arranged: In Sec. 2, we give a brief summary of the NU method. In Sec. 3, we apply NU method for solving the hyper-radial

Schrödinger equation. In Sec. 4, the results are discussed. In Sec. 5, the summary and conclusion are written.

## 2 –Formalism

The Nikiforov-Uvarov method is one of the strong techniques to find the analytic solutions of a generalized second-order linear differential equation with special orthogonal functions of the type [37],

$$\psi''(z) + \frac{\tilde{\tau}(z)}{\sigma(z)}\psi'(z) + \frac{\tilde{\sigma}(z)}{\sigma^2(z)}\psi(z) = 0, \qquad (1)$$

where $\sigma(z)$ and $\tilde{\sigma}(z)$ are polynomials of maximum second degree and $\tilde{\tau}(z)$ is a polynomial of maximum first degree. In order to solve Eq. (1), we write $\psi(z)$ as

$$\psi(z) = \chi(z)\phi(z). \qquad (2)$$

Substitution of Eq. (2) into Eq. (1), the hypergeometric type is obtained as

$$\sigma(z)\chi''(z) + \tau(z)\chi'(z) + \lambda\chi(z) = 0. \qquad (3)$$

In Eq. (2), the wave function $\phi(z)$ is defined

$$\frac{\phi'(z)}{\phi} = \frac{\pi(z)}{\sigma(z)}, \qquad (4)$$

with $\pi(z)$ is most first-order polynomials. Also, the hypergeometric-type functions in Eq. (3) for a fixed integer $n$ is given by the Rodrigue relation as

$$\chi_n = \frac{B_n}{\rho_n}\frac{d^n\left[\sigma^n(z)\rho(z)\right]}{dz^n} \qquad (5)$$

where $B_n$ is the normalization constant and the weight function $\rho(z)$ must satisfy the condition

$$\frac{d^n}{dz}[\sigma(z)\rho(z)] = \tau(z)\rho(z), \qquad (6)$$

with

$$\tau(z) = \tilde{\tau}(z) + 2\pi(z). \tag{7}$$

The function $\pi(z)$ and the parameter $\lambda$ required for the NU method are then defined

$$\pi(z) = \frac{\sigma'(z) - \tilde{\tau}(z)}{2} \pm \sqrt{\left(\frac{\sigma'(z) - \tilde{\tau}(z)}{2}\right)^2 - \tilde{\sigma}(z) + k\sigma(z)} \tag{8}$$

$$\lambda = k + \pi'(z) \tag{9}$$

The values of k in the square-root of Eq. (8) is possible to calculate if the expression under the square root are square of expression. This is possible if its discriminate is zero.

Therefore, the new eigenvalue equation becomes

$$\lambda = \lambda_n = -n\tau' - \frac{n(n-1)}{2}\sigma'', n = 0, 1, 2... \tag{10}$$

## 3 - Solution of the hyper-radial Schrödinger equation in the anisotropic medium

In the nonrelativistic quark model, baryons are formed by three constituent quarks. In Ref. [30], the authors show that the potential of quark-quark system is the half of that of corresponding quark-antiquark system $V_{qq} = \frac{1}{2}V_{q\bar{q}}$. So, the Cornell potential is written as follows

$$V_{qq} = \sum_{i,j=1, i \neq j}^{3} \frac{1}{2}(\sigma r_{ij} - \frac{\alpha}{r_{ij}}), \tag{11}$$

where $r_{ij} = |\vec{r}_i - \vec{r}_j|$ is the distance between the qq pair. The parameter $\sigma$ = 0.184 GeV$^2$ as in Refs. [3, 39]. The real Cornell potential of the quark-antiquark interaction in an anisotropic hot medium in which the quark and antiquark pairs aligned to the direction of anisotropy, yields as [23],

$$V_{q\bar{q}}(r) = \left(\frac{2\sigma}{m_D^2} - \alpha\right)\frac{e^{-m_D r}}{r} - \frac{2\sigma}{m_D^2 r} + \frac{2\sigma}{m_D} - \alpha m_D$$
$$+ \xi\left[\frac{4\sigma}{m_D^2 r}e^{-m_D r}\left(\frac{2(e^{-m_D r}-1)}{(m_D r)^2} - \frac{(e^{-m_D r}+2)}{3} - \frac{2}{m_D r} - \frac{m_D r}{12}\right) - \frac{\alpha m_D}{m_D r}e^{-m_D r}\left(\frac{2(e^{-m_D r}-1)}{(m_D r)^2} - \frac{2}{m_D r} - \frac{m_D r}{6} - 1\right)\right]$$
(12)

In the present work, the Debye mass $D(T, u_b)$ is given as in Ref. [11] and references therein by

$$D(T, u_b) = g(T)T\sqrt{\frac{N_c}{3} + \frac{N_f}{6} + \frac{N_f}{2\pi^2}\left(\frac{u_b}{3T}\right)^2}$$
(13)

where, $g(T)$ is the coupling constant as defined in Refs. [22, 39], $u_b$ is the baryon chemical potential, $N_f$ is number of flavours, and $N_c$ is number of colors. $\xi$ is an anisotropic parameter. This potential reduces to Cornell potential at zero temperature at $m_D = \xi = 0$ as in Ref. [23]. From Ref. [30] that gave the relation between the quark-antiquark system and quark-quark system, we can write the potential of quark-quark system as follows

$$V_{qqq}(r) = \frac{1}{2}\sum_{i,j=1,i\neq j}^{3}\left(\frac{2\sigma}{m_D^2} - \alpha\right)\frac{e^{-m_D r_{ij}}}{r_{ij}} - \frac{2\sigma}{m_D^2 r_{ij}} + \frac{2\sigma}{m_D} - \alpha m_D$$
$$+ \xi\left[\frac{4\sigma}{m_D^2 r_{ij}}e^{-m_D r_{ij}}\left(\frac{2(e^{-m_D r_{ij}}-1)}{(m_D r)^2} - \frac{(e^{-m_D r_{ij}}+2)}{3} - \frac{2}{m_D r_{ij}} - \frac{m_D r_{ij}}{12}\right) - \frac{\alpha m_D}{m_D r_{ij}}\left(\frac{2(e^{-m_D r_{ij}}-1)}{(m_D r_{ij})^2} - \frac{2}{m_D r_{ij}} - \frac{m_D r_{ij}}{6} - 1\right)\right]$$
(14)

The three body system can be well described by two Jacobi vectors $\vec{\rho}$ and $\vec{\lambda}_1$ where

$$\vec{\rho} = \frac{1}{\sqrt{2}}(\vec{r_1} - \vec{r_2}) \text{ and } \vec{\lambda}_1 = \frac{1}{\sqrt{6}}(\vec{r_1} + \vec{r_2} - 2\vec{r_3}).$$
(15)

In the hyperspherical coordinates, the hyper-radius $x$ is introduced via

$$x = \sqrt{\rho^2 + \lambda_1^2},$$
(16)

As in Refs. [30, 41], the potential in Eq. (14) is replaced in the hyperspherical coordinates for quark-quark system by

$$V_{qqq}(x) = \frac{1}{2}\left\{\left(\frac{2\sigma}{m_D^2}-\alpha\right)\frac{e^{-m_D x}}{x} - \frac{2\sigma}{m_D^2 x} + \frac{2\sigma}{m_D} - \alpha m_D \right.$$

$$+\zeta\left[\frac{4\sigma}{m_D^2 x}e^{-m_D x}\left(\frac{2(e^{-m_D x}-1)}{(m_D r)^2} - \frac{(e^{-m_D x}+2)}{3} - \frac{2}{m_D x} - \frac{m_D x}{12}\right) - \frac{\alpha m_D}{m_D x}\left(\frac{2(e^{-m_D x}-1)}{(m_D x)^2} - \frac{2}{m_D x} - \frac{m_D x}{6} - 1\right)\right]\right\} .$$

(17)

The hyperspherical Schrödinger equation with effective Cornell potential takes the form as in Ref. [40],

$$\left\{\frac{d^2}{dx^2} + \frac{5}{x}\frac{d}{dx} - \frac{\gamma(\gamma+4)}{x^2} + 2\mu\left(E_{n,\gamma} - V^{eff}(x)\right)\right\}R_{n,\gamma}(x) = 0 \quad (18)$$

where $\mu, \gamma, E_{n,\gamma}, R_{n,\gamma}(x)$ and $V^{eff}(x)$ are the reduced mass of the baryon, grand angular momentum quantum number, energy of the baryon, the hyper-radial wave function, and the effective potential. The reduced mass is defined as in Ref. [40].

$$\mu = \frac{2m_\rho m_{\lambda_1}}{m_\rho + m_{\lambda_1}} \quad (19)$$

where

$$m_\rho = \frac{2m_1 m_2}{m_1 + m_2}, \quad m_{\lambda_1} = \frac{3m_1(m_1+m_2)}{(m_1+m_2+m_3)} \quad (20)$$

$m_1, m_2,$ and $m_3$ are masses of the first, second, and third of constituents of each baryon. Introducing the gauge transformation

$$R_{n,\gamma}(x) = \varphi_{n,\gamma}(x) x^{-\frac{5}{2}} \quad (21)$$

The Eq. (18) takes the following form

$$\frac{d^2\varphi_{n,\gamma}(x)}{dx^2} + 2\mu\left[E_{n,\gamma} - V^{eff}(x) - \frac{\frac{15}{4} + \gamma(\gamma+4)}{2\mu x^2}\right]\varphi_{n,\gamma}(x) = 0, \tag{22}$$

where, the effective potential for the quark-quark system is defined as,

$$V^{eff}(x) = V(x) - V(x \to \infty) = \frac{1}{2}\left\{\left(\frac{2\sigma}{m_D^2} - \alpha\right)\frac{e^{-m_D x}}{x} - \frac{2\sigma}{m_D^2 x}\right.$$

$$+\varsigma\left[\frac{4\sigma}{m_D^2 x}e^{-m_D x}\left(\frac{2(e^{-m_D x}-1)}{(m_D r)^2} - \frac{(e^{-m_D x}+2)}{3} - \frac{2}{m_D x} - \frac{m_D x}{12}\right) - \frac{\alpha m_D}{m_D x}\left(\frac{2(e^{-m_D x}-1)}{(m_D x)^2} - \frac{2}{m_D x} - \frac{m_D x}{6} - 1\right)\right]\}$$

(23)

Substituting Eq. (23) into Eq. (22) after making expansion for the exponential function up to second order correction as $e^{-m_D x} \approx 1 - m_D x + \frac{m_D^2 x^2}{2}$ yields,

$$\frac{d^2\varphi_{n,\gamma}(x)}{dx^2} + \left[2\mu E_{n,\gamma} - \mu\left(\frac{a}{x^2} + \frac{b}{x} + c + dx + ex^2 + fx^3\right) - \frac{\frac{15}{4} + \gamma(\gamma+4)}{x^2}\right]\varphi_{n,\gamma}(x) = 0, \tag{24}$$

where,

$$a = \varsigma\left(\frac{4\alpha}{m_D} - \frac{16\sigma}{m_D^3}\right), b = -\alpha + \varsigma\left(\frac{16\sigma}{m_D^2} - 4\alpha\right),$$

$$c = \varsigma\left[-\frac{2\sigma}{m_D} + \frac{13}{6}\alpha m_D\right], d = \sigma - \frac{\alpha m_D^2}{2} + \varsigma\left(-\frac{5\sigma}{3} - \frac{1}{6}\alpha m_D\right),$$

$$e = \varsigma\left[\frac{7\sigma m_D}{6} + \frac{1}{12}\alpha m_D^3\right]$$

$$f = -\varsigma\left(\frac{\sigma m_D^2}{3}\right)$$

(25)

Using the change of variable, $s = \frac{1}{x}$, equation (24) becomes,

$$\frac{d^2\varphi_{n,\gamma}(s)}{ds^2}+\frac{2s}{s^2}\frac{d\varphi_{n,\gamma}(s)}{ds}+\frac{1}{s^4}\left(2\mu E_{n,\gamma}-\mu c-\frac{\mu d}{s}-\frac{\mu e}{s^2}-\frac{\mu f}{s^3}-\mu b s-(\gamma(\gamma+4)+\frac{15}{4}+\mu a)s^2\right)\varphi_{n,\gamma}(s)=0.$$
(26)

In order to solve equation (26), we make the following approximation scheme on the terms $s$ as in Ref. [11] with assumption that $x_0$ is the characteristic radius of the baryon that defined in Ref. [42]. Setting $z=(s-\delta)$ and expanding in a power series around $z=0$, where $\delta=\frac{1}{x_0}$, we get

$$\frac{1}{s}=\frac{1}{(\delta+z)}=\frac{1}{\delta}(1+\frac{z}{\delta})^{-1}\approx\frac{1}{\delta^3}\left(3\delta^2-3s\delta+s^2\right),$$

$$\frac{1}{s^2}=\frac{1}{(\delta+z)^2}=\frac{1}{\delta^2}(1+\frac{z}{\delta})^{-2}\approx\frac{1}{\delta^4}\left(6\delta^2-8s\delta+3s^2\right),$$

$$\frac{1}{s^3}=\frac{1}{(\delta+z)^3}=\frac{1}{\delta^3}(1+\frac{z}{\delta})^{-3}\approx\frac{1}{\delta^5}\left(10\delta^2-15s\delta+6s^2\right),$$

(27)

Substituting Eq. (27) into Eq. (26) yields,

$$\frac{d^2\varphi_{n,\gamma}(s)}{ds^2}+\frac{2s}{s^2}\frac{d\varphi_{n,\gamma}(s)}{ds}+\frac{2}{s^4}\left\{-x_1 s^2+x_2 s-x_3\right\}\varphi_{n,\gamma}(x)=0 \qquad (28)$$

where,

$$x_1=\mu\left(\frac{d}{2\delta^3}+\frac{3e}{2\delta^4}+\frac{3f}{\delta^5}+\frac{a}{2}+\frac{(\gamma(\gamma+4)+\frac{15}{4})}{2\mu}\right)$$

$$x_2=-\mu\left(\frac{3d}{2\delta^2}+\frac{4e}{\delta^3}+\frac{15f}{2\delta^4}-\frac{b}{2}\right),$$

$$x_3=\mu\left(-E_{n,\gamma}+\frac{c}{2}+\frac{3d}{2\delta}+\frac{3e}{\delta^2}+\frac{5f}{\delta^3}\right),$$

(29)

Now comparing Eq. (28) with Eq. (1), we obtain the following relation which is amendable for NU method,

$$\tilde{\tau}(s) = 2s, \sigma(s) = s^2,$$
$$\tilde{\sigma}(s) = 2(-x_1 s^2 + x_2 s - x_3) \tag{30}$$

Now using Eq. (9), we obtain

$$\pi(s) = \pm\sqrt{(k + 2x_1)s^2 - 2x_2 s + 2x_3} \tag{31}$$

The values of $k$ in the square-root of Eq. (31) is possible to calculate if the expression under the square root are square of expression. This is possible if its discriminate is zero. Thus,

$$\pi(s) = \pm \frac{1}{\sqrt{2x_3}}(x_2 s - 2x_3) \tag{32}$$

with $k$-values given as,

$$k = \left(\frac{x_2^2}{2x_3} - 2x_1\right) \tag{33}$$

Using Eq. (7), we obtain

$$\tau(s) = 2s - \frac{2}{\sqrt{2x_3}}(x_2 s - 2x_3) \tag{34}$$

From equations (9) and (10), we obtain

$$\lambda = \left(\frac{x_2^2}{2x_3} - 2x_1\right) - \frac{x_2}{\sqrt{2x_3}}, \tag{35}$$

$$\lambda_n = -n - n^2 + \frac{2nx_2}{\sqrt{2x_3}} \tag{36}$$

By comparing these expressions, $\lambda = \lambda_n$, we obtain the energy expression for the heavy baryon as

$$\frac{x_2^2}{2x_3} - 2x_1 - \frac{x_2}{\sqrt{2x_3}} = \frac{2nx_2}{\sqrt{x_3}} - n^2 - n \tag{37}$$

let $y = \frac{x_2}{\sqrt{2x_3}}$ and this result to the following quadratic equation,

$$y^2 - (2n+1)y + n(n+1) - 2x_1 = 0 \tag{38}$$

Solving equation (38) completely, we obtain the energy spectrum for each baryon as,

$$E_{n,\gamma} = \frac{c}{2} + \frac{3d}{2\delta} + \frac{3e}{\delta^2} + \frac{5f}{\delta^3}$$

$$- \frac{\mu\left(\frac{3d}{2\delta^2} + \frac{4e}{\delta^3} + \frac{15f}{2\delta^4} - \frac{b}{2}\right)^2}{\left[2n+1 \pm \sqrt{1 + 8\mu\left(\frac{d}{2\delta^3} + \frac{3e}{2\delta^4} + \frac{3f}{\delta^5} + \frac{a}{2} + \frac{(\gamma(\gamma+4)+\frac{15}{4})}{2\mu}\right)}\right]^2} \tag{39}$$

The corresponding wave function can be obtain as,

$$R_{n,\gamma}(x) = N_{n\gamma} x^{-\frac{2\chi_2 + 5\sqrt{2\chi_1}}{2\sqrt{2\chi_1}}} e^{\sqrt{2\chi_1}x} \left(\frac{d}{dx}\right)^n \left(x^{-2n+\frac{2\chi_2}{\sqrt{2\chi_1}}} e^{-\sqrt{2\chi_1}x}\right) \tag{40}$$

where $N_{n\gamma}$ is the normalization constant.

## 3-Discussion of Results

In this section, we discuss the dissociation of baryons such as nucleon and some heavy baryon such as $\Lambda_c^+, \Xi_c^+, \Omega_c^0$ in the hot and dense media. In the first, we discuss the behavior of real potential which plays an essential role in the present analysis. We know that the nucleon consistent of three quark masses from two up quark and one down quark which has $m_u = m_d = 0.336$ GeV as in Ref. [40]. We consider the weakly anisotropy parameter ($\xi \leq 1$) at finite temperature and finite baryonic chemical potential in QCD plasma as in Ref. [23].

In Fig. (1), the effective potential is given by Eq. (23) is plotted as a function of $r$ at different values of temperatures above the critical temperature $T_c$ = 170 MeV. We note that the potential has negative values for all values of r which means the potential is an attractive potential. The confinement part of potential disappears that temperature is taken above the critical temperature. Also, we note that the potential increases slightly by increasing an anisotropic parameter at a small distance of r. In Ref. [29], in the isotropic medium, medium modification to the linear term remains positive up to 2-3 $T_c$, making the potential less attractive. In contrast, in the anisotropic case medium modification to the linear term becomes negative and the overall full potential becomes more attractive. In Ref. [24], the potential is attractive for all values of $r$ and the potential is sensitive to $\xi$ when $\xi$ is greater than 1. In the present study, we restricted the present calculations to the weakly anisotropic parameter $\xi \leq 1$ as in Refs. [23, 29]. In Fig. (2), we note that the potential is more attractive by increasing anisotropic parameter.

In Figs. (3 and 4), the potential is plotted in the two cases. In the first case, the effect of anisotropic medium is not considered, we see that the potential strongly increases by increasing temperature and slightly increases by increasing baryon chemical potential which means that the hot medium is more affected than a dense medium. This indicates the dissociation of baryon will be strong in the hot medium. In the second case, the anisotropic medium is considered, we note that the potential is more attractive.

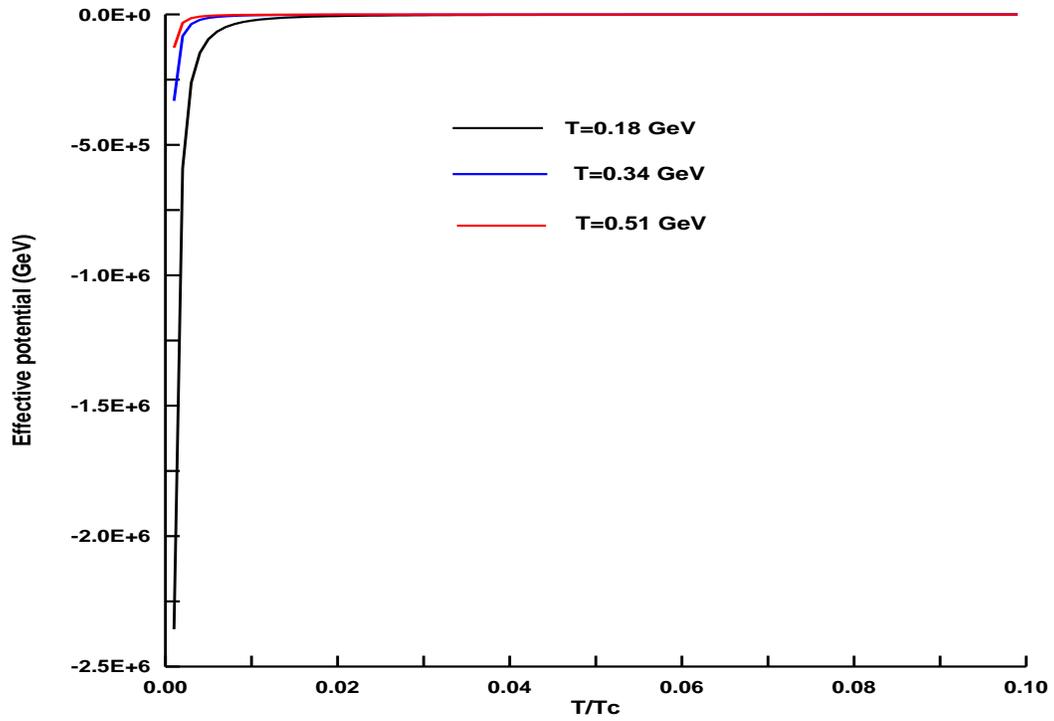

Fig. (1): The effective potential is plotted as a function of distance r for different values of temperature at $\xi = 0.12$

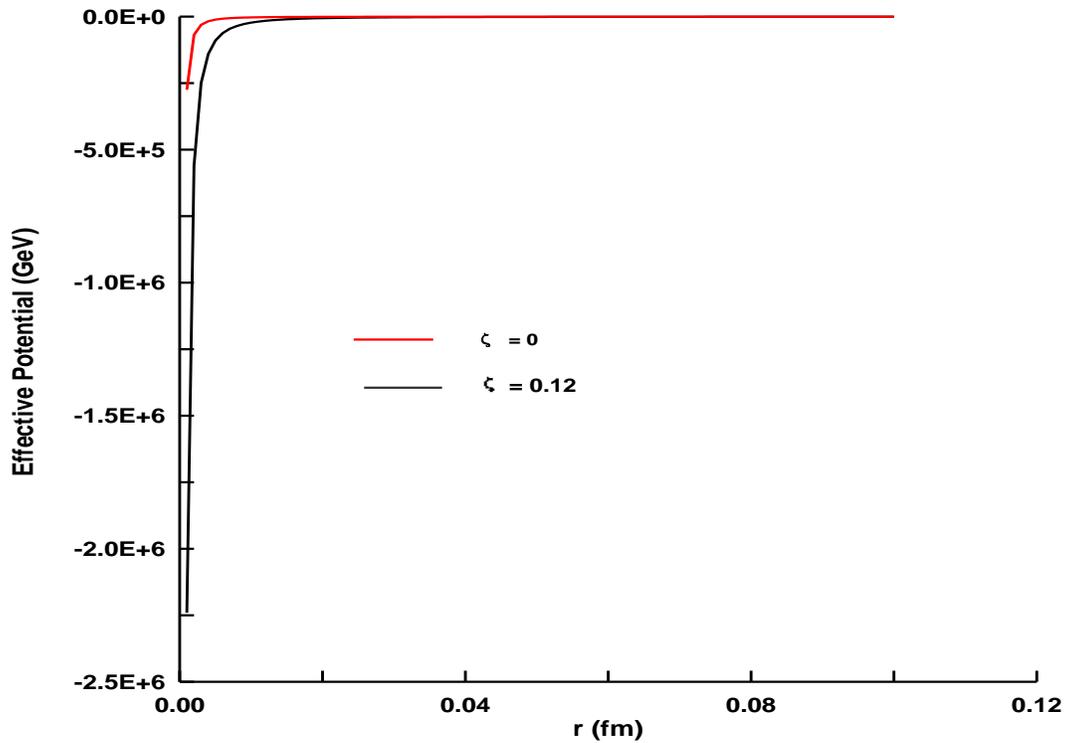

Fig. (2): The effective potential is plotted as a function of distance r for two values of $\xi$

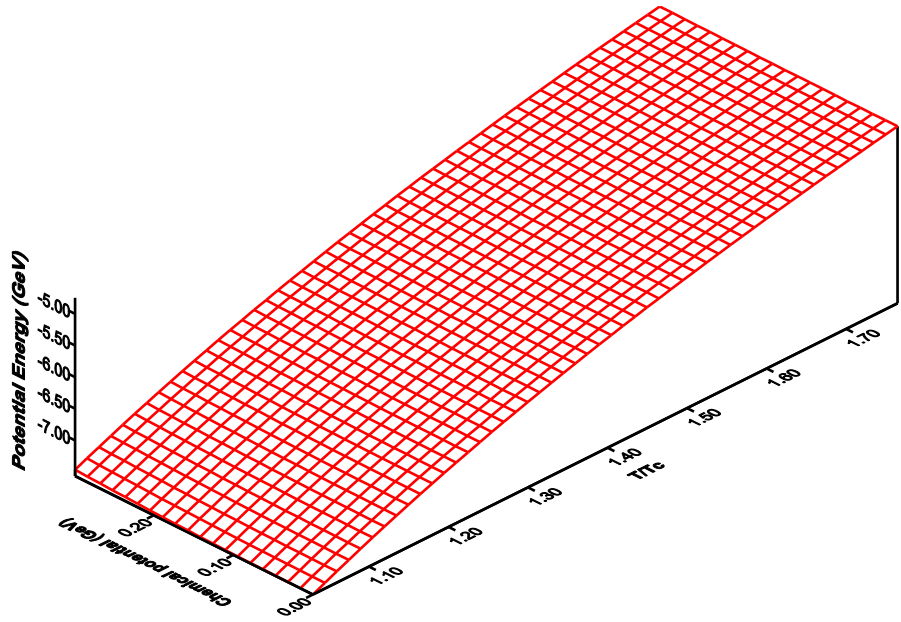

Fig. (3): The potential is plotted as a function of temperature and baryonic chemical potential in an isotropic medium

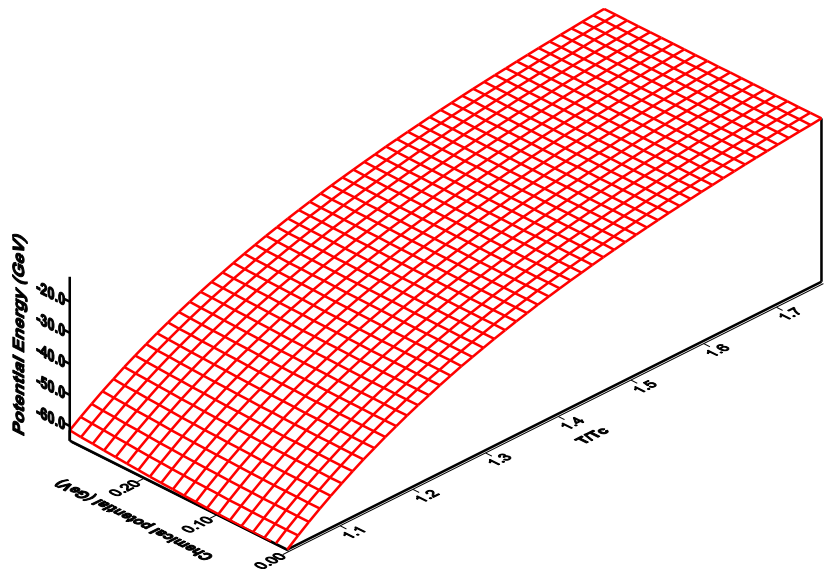

Fig. (4): The potential is plotted as a function of temperature and baryonic chemical potential in an anisotropic medium

## 3.1 Binding energy of Nucleon state

In the previous section, we obtained the exact solution of hyper-Schrodinger equation using NU method. We obtained binding energy that depends on temperature, chemical potential, and anisotropic parameter. In this subsection, we investigate the behavior of nucleon with these parameters and compare with other studies.

In Fig. (5), the binding energy of nucleon is plotted as a function of temperature ratio where the critical temperature equal to $T_c$=170 MeV as in Ref. [29]. In Fig. (5), we consider a hot medium only that means the baryon chemical potential is zero. We note that the binding energy decreases with increasing temperature. This is indicated that the binding energy of nucleon dissociates by increasing temperature after the critical temperature. This finding is a good agreement with Ref. [30], in which the radial Schrodinger is solved using Gaussian expansion method in an isotropic medium. By increasing anisotropic parameter, we see that the binding energy of nucleon shifts to higher values that indicate the nucleon state is more bound in the anisotropic medium. This is finding is not considered before in the previous studies that are concentrated on dissociation of quarkonium. Thus, this is finding is an agreement with other studies in the qurakonium states such as in Refs. [23, 29].

In Fig. (6), we consider a finite baryon chemical potential equal to 300 MeV. Similar behavior is obtained in comparison with Fig. (5) and we see that the binding energy slightly shifts to lower values by increasing baryon chemical potential in hot medium above the critical temperature. Therefore, the bound state is less bound in the hot and dense media. This indicates that the screening of the Coulomb and string contributions are less accentuated and hence the nucleon binding energy state become larger than in the case of isotropic medium.

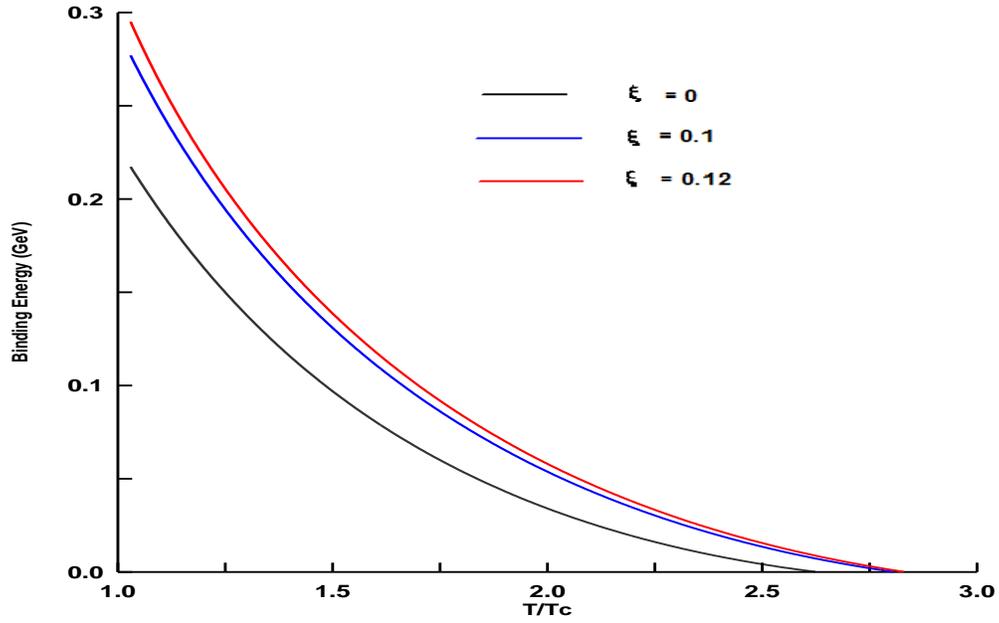

**Fig. (5)**. The binding energy of nucleon is plotted as a function of temperature at zero chemical potential

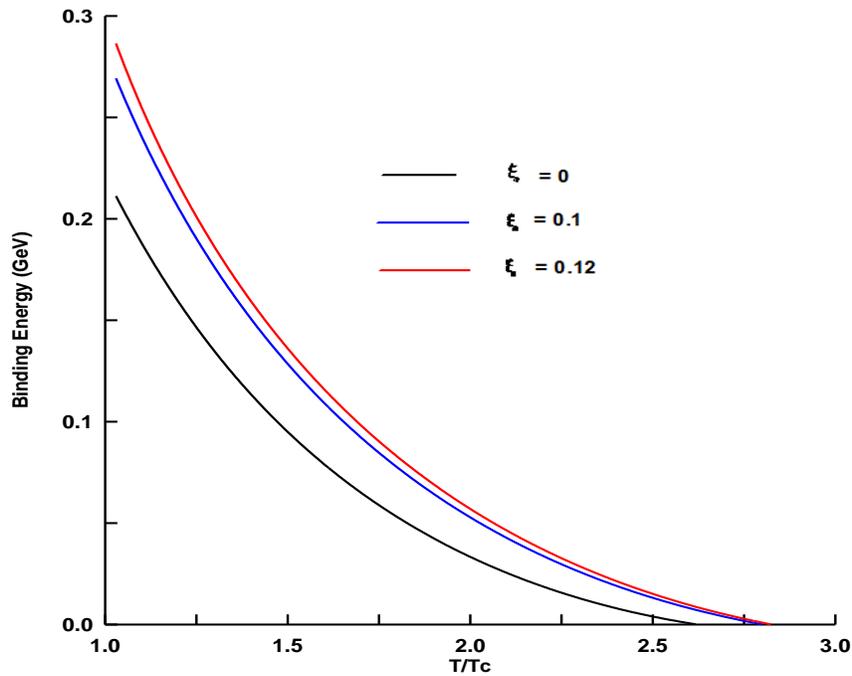

**Fig. (6)**. The binding energy of nucleon is plotted as a function of temperature at finite chemical potential = 0.3 GeV

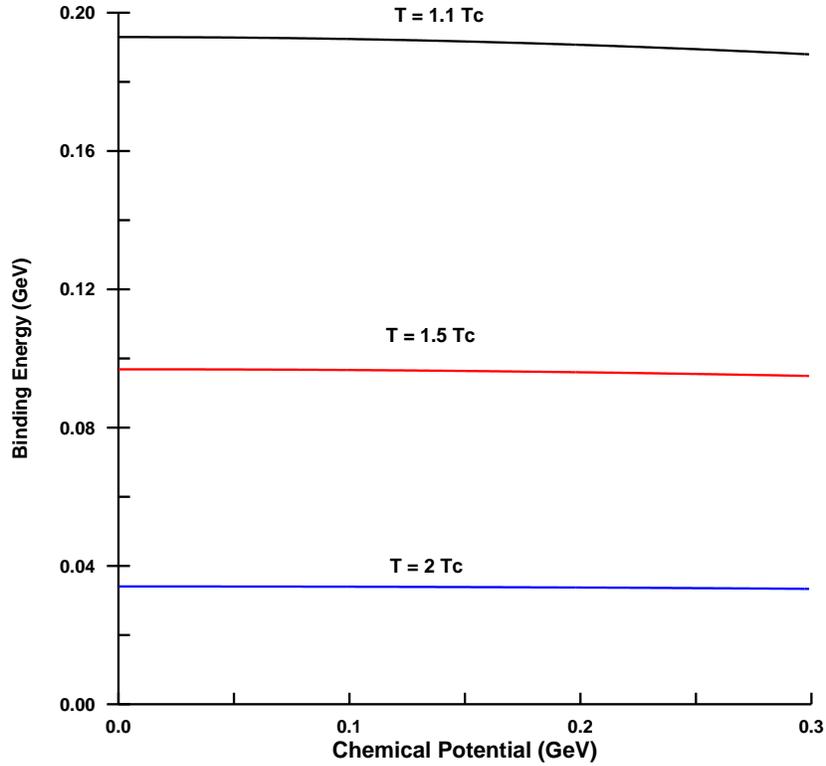

**Fig. (7).** The binding energy of nucleon is plotted as a function of baryon chemical potential for three values of temperatures above critical temperature in the isotropic medium

In Fig. (7), the binding energy slightly decreases with increasing baryon of chemical potential for different values of temperature in the isotropic medium. This means that the binding energy is a little sensitive for changing of baryon chemical potential in the hot medium.

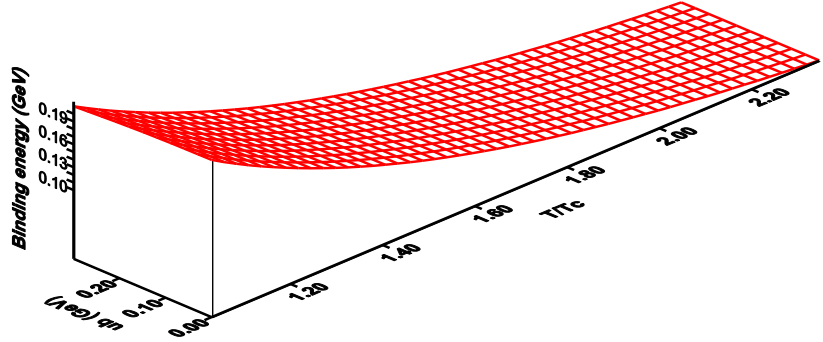

**Fig. (8)**. The binding energy of is plotted as a function of temperature and chemical potential in the isotropic medium

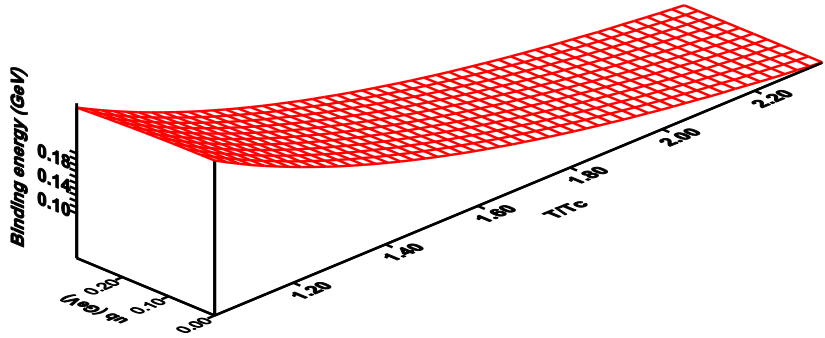

**Fig. (9)**. The binding energy of nucleon is plotted as a function of temperature and chemical potential in the anisotropic medium

In Figs. (8 and 9), the binding energy is plotted in 3D for the isotropic medium ( $\xi = 0$ ) and anisotropic medium ( $\xi = 0.12$), respectively. We note that the binding energy decreases with increasing temperature for any value of the baryonic chemical potential. Also, we note that the binding energy slightly decreases with increasing baryon chemical potential. Moreover, we note that the binding energy is high in the anisotropic medium. Therefore, the bound state is enhanced in an anisotropic medium.

## 3.2- Dissociation of baryon at finite temperature and chemical potential

The dissociation of a two-body bound state in a thermal medium can be given qualitatively: when the binding energy of a resonance state drops below the mean-thermal energy of quarkonium i. e. the state becomes weakly bound. In Ref. [23] and references therein, the authors concluded that one need not to reach the binding energy ($E_b$) to be zero for the dissociation but a weaker condition is assumed $E_b \leq T$ which causes a state to be weakly bound. In fact, when $E_b \simeq T$, the resonances have been broadened due to direct thermal activation. So the dissociation of the bound states may be expected to occur roughly around $E_b \simeq T$. In the present analysis, we investigate the dissociation of nucleon when an anisotropic medium is considered in hot medium or hot and dense medium when the dissociation of the bound states may be expected to occur roughly around $E_b \simeq T$ as discussed above.

In Table (1), we note that the dissociation of temperature increases with increasing anisotropic parameter due to the binding energy is more bound in anisotropic medium. By increasing baryon chemical potential, we note that the dissociation of temperature is little sensitive for changing baryon chemical potential. In Ref. [30], the authors found that the dissociation temperature of nucleon is accrued above critical temperature about 1.17 $T_c$. They considered isotropic medium at zero baryon chemical potential and calculated the dissociation temperature by using a different technique. Therefore, the agreement is obtained in comparison with Ref. [30] that $T_D$ = 1.1 GeV in the present work. We have two advantages that the present study is extended an anistropoc medium in the hot medium and also the effect of finite baryon chemical is considered. We note that dissociation temperature $(T_D)$ increases by increasing anisotropic parameter $\xi$ due to the binding energy is more bound in an anisotropic medium. By increasing baryon chemical potential up to 300 MeV, we note that the dissociation temperature slightly decreases in comparison with its value at zero baryon chemical potential. Thus, the nucleon state is less bound in hot and dense media.

In Fig. (10), the binding energy of heavy baryons $\Lambda_c^+(cud)$, $\Xi_c^+(cus)$, and $\Omega_c^0(css)$ is plotted as a function of temperature ratio at zero baryon chemical potential. We calculate the binding energy of each baryon by using Eq. (39) in which the reduced mass is calculated as in Eq. (19) for each baryon such as $\Lambda_c^+(cud)$, we took $m_1 = m_u = 0.336 \, \text{GeV}$, $m_2 = m_d = 0.336 \, \text{GeV}$ and $m_3 = m_c = 1.55 \, \text{GeV}$ as in Ref. [40]. We note all states of baryon decrease by increasing temperature that means every baryon melts in an isotropic medium. In the anisotropic medium at $\xi = 0.12$, we note that the binding of each heavy baryon is shifted to higher values which mean the baryon is tight in this medium as seen in Fig. (11).

In Table (2), we note that the dissociation of temperature increases with increasing anisotropic parameter for each baryon. $\Omega_c^0(css)$ gives a high value in comparison with other heavy baryons due to its binding energy is more tight in the isotropic and anisotropic media.

**Table (1)**: Dissociation temperature $(T_D)$ for the Nucleon

|  | $\zeta = 0$ | $\zeta = 0.1$ | $\zeta = 0.12$ |
|---|---|---|---|
| $T_D \mid_{\mu_b=0}$ | 1.11 $T_c$ | 1.21 $T_c$ | 1.235 $T_c$ |
| $T_D \mid_{\mu_b=0.3}$ | 1.10 $T_c$ | 1.20 $T_c$ | 1.229 $T_c$ |

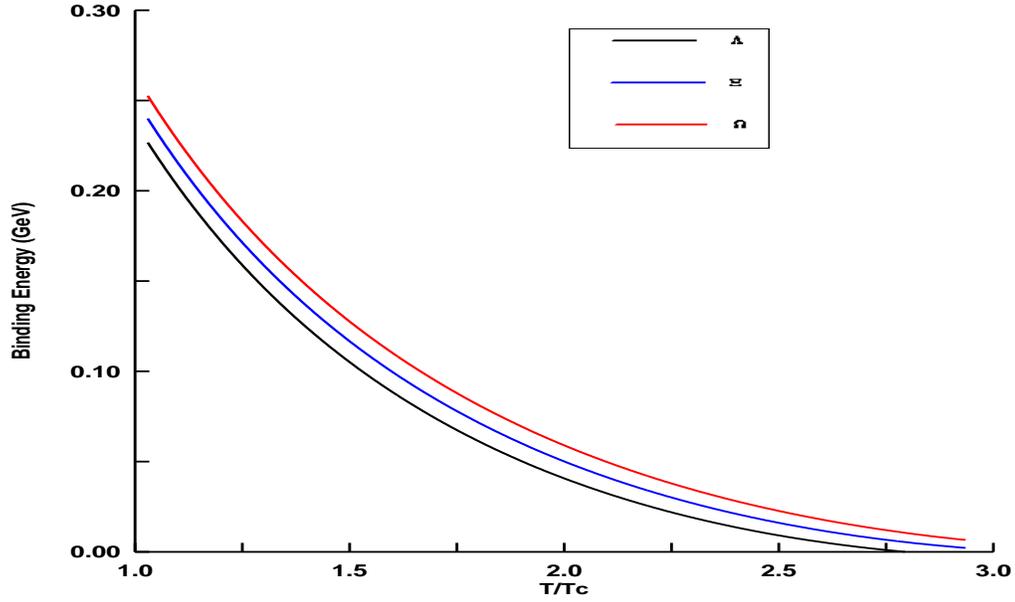

**Fig. 10**. The binding energy of heavy baryon $\wedge_c^+(cud)$, $\Xi_c^+(cus)$, and $\Omega_c^0(css)$ at zero baryon chemical in the isotropic medium $\zeta = 0$

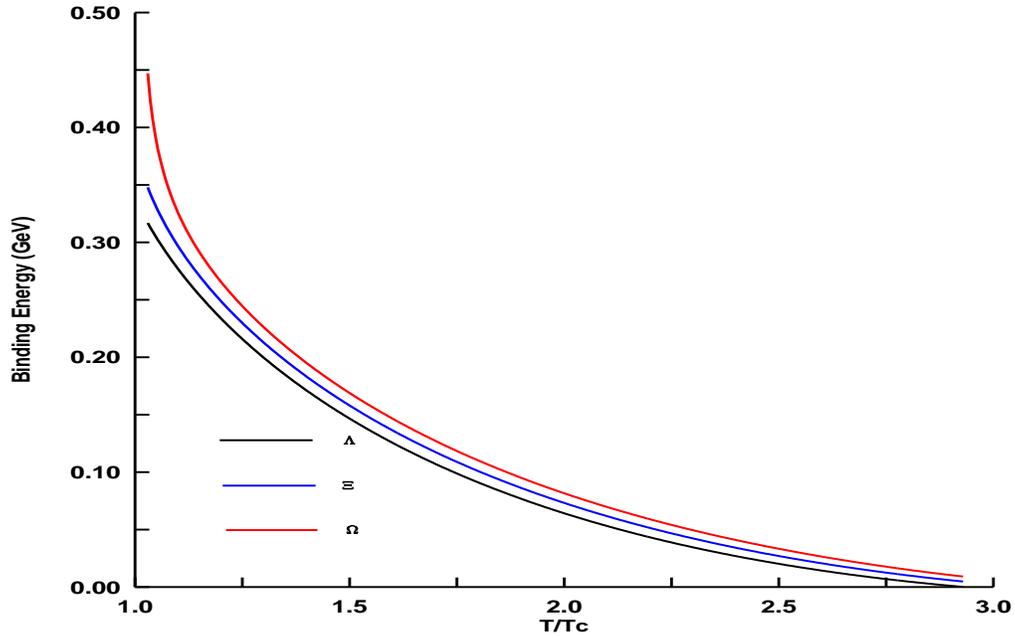

**Fig. 11**. The binding energy of heavy baryon $\wedge_c^+(cud)$, $\Xi_c^+(cus)$, and $\Omega_c^0(css)$ at zero baryon chemical in the anisotropic medium $\zeta = 0.12$

**Table (2)**: Dissociation temperature $(T_D)$ of heavy-baryon

| **Heavy-baryon** | $\zeta = 0$ and $\mu_b = 0$ | $\zeta = 0.12$ and $\mu_b = 0$ |
|---|---|---|
| $\Lambda_c^+(cud)$ | 1.129 $T_c$ | 1.2588 $T_c$ |
| $\Xi_c^+(cus)$ | 1.1588 $T_c$ | 1.2823 $T_c$ |
| $\Omega_c^0(css)$ | 1.1824 $T_c$ | 1.312 $T_c$ |

## Summary and Conclusion

In this paper, the dissociation temperature for baryons such as N(uud), $\Lambda_c^+(cud)$, $\Xi_c^+(cus)$, and $\Omega_c^0(css)$ have been calculated using hyper-Schrödinger equation, in which the effective real quark potential depends on the finite temperature and baryon chemical potential. The Nikiforov-Uvarov method is a strong technique to obtain the analytic solution of hyper-radial Schrödinger equation. The energy eigenvalues and corresponding wave functions are obtained. We have applied the present results in the isotropic and anisotropic media. We found that the binding energy of nucleon and heavy baryon strongly decreases with increasing temperature and slightly decreases with increasing baryon chemical potential in the isotropic and anisotropic media. In anisotropic medium, we found that the binding energy for every baryon is more bound in comparison with their values in the isotropic medium. The dissociation temperature of nucleon is investigated. We found that the nucleon melts above critical temperature about $1.1 T_c$ in an isotropic medium. The dissociation temperature increases in the anisotropic medium due to the binding energy is more bound in the anisotropic medium. The baryon chemical potential is slightly affected on the dissociation temperature. In Ref. [30], authors found that the nucleon melts above critical temperature in an isotropic medium. The present finding is a good agreement with the result of Ref. [30] that the nucleon melts above critical temperature about 1.17 $T_c$. The present result show that the dissociation temperature of heavy baryon melt above critical temperature in the isotropic and anisotropic media. The baryon of chemical potential is slightly acted on the binding energy and dissociation of temperature in the isotropic and anisotropic media. For our best knowledge, this is a new treatment for some baryons in an anisotropic medium in hot QCD medium when the baryon chemical potential is considered. Thus, the present work provides a good description for dissociation of baryon in the isotropic and anisotropic media to analysis the baryon rich quark gluon plasma which is expected at FAIR energies will be created. We hope to investigate in a future work the dissociation temperature using the decay width of baryon bound states from the imaginary part of the potential in the dense medium.